\documentclass[aps,twocolumn]{revtex4}
\usepackage{latexsym}
\usepackage{amsfonts}
\usepackage{graphicx} 
\usepackage{epsfig}
\begin{document}


\title{ Relation Between the Widom line and the Strong-Fragile Dynamic
Crossover in Systems with a Liquid-Liquid Phase Transition}

\author{L. Xu$^{\ast}$, P. Kumar$^{\ast}$,
S. V. Buldyrev$^{\dagger\ast}$, S.-H. Chen$^{\ddagger}$,
P. H. Poole$^{\S}$, F. Sciortino$^{\P}$, H. E. Stanley$^{\ast}$ } 

\bigskip
\bigskip

\affiliation{ $^{\ast}$Center for Polymer Studies and Department of Physics,
Boston University,~Boston, MA 02215 USA\\ $^{\dagger}$Department of
Physics,~Yeshiva University, 500 West 185th Street,~New York, NY 10033 USA \\
$^{\ddagger}$Nuclear Science and Engineering Department,~Massachusetts
Institute of Technology, Cambridge, MA 02139 USA\\ $^{\S}$Department of
Physics, St. Francis Xavier University, Antigonish, Nova Scotia B2G 2W5,
Canada \\ $^{\P}$Dipartimento di Fisica and Istituto Nazionale Fiscia della
Materia Unita' di Ricerca: Complex Dynamics in Structured Systems,
Universita' di Roma ``La Sapienza'' -- Piazzale Aldo Moro 2, I-00185, Roma,
Italy }


\begin{abstract}
We investigate, for two water models displaying a liquid-liquid critical
point, the relation between changes in dynamic and thermodynamic
anomalies arising from the presence of the liquid-liquid critical
point. We find a correlation between the dynamic fragility transition
and the locus of specific heat maxima $C_P^{\rm max}$ (``Widom line'')
emanating from the critical point. Our findings are consistent with a
possible relation between the previously hypothesized liquid-liquid
phase transition and the transition in the dynamics recently observed in
neutron scattering experiments on confined water. More generally, we
argue that this connection between $C_P^{\rm max}$ and dynamic crossover
is not limited to the case of water, a hydrogen bond network forming
liquid, but is a more general feature of crossing the Widom
line. Specifically, we also study the Jagla potential, a
spherically-symmetric two-scale potential known to possess a
liquid-liquid critical point, in which the competition between two
liquid structures is generated by repulsive and attractive ramp
interactions.
\end{abstract}
\bigskip

\maketitle
 \section{Introduction} 

\begin{figure}
\includegraphics[width=8.5cm]{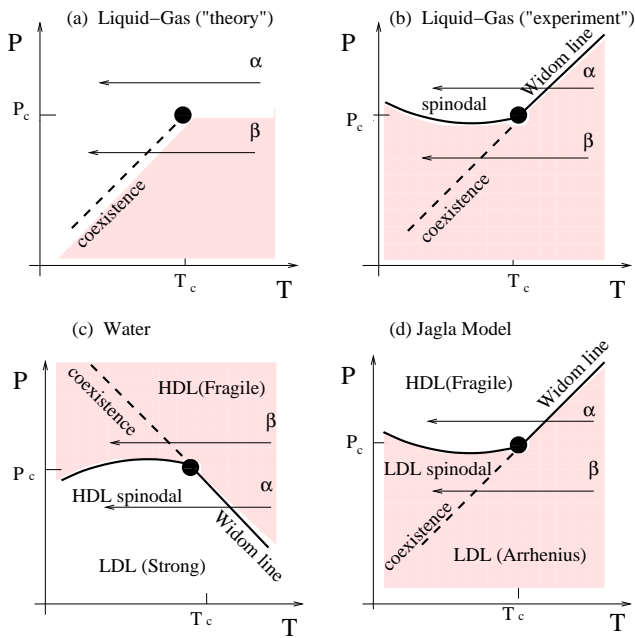}
\caption{ Schematic phase diagram for the systems discussed in this
paper. (a) The critical region associated with a liquid-gas critical
point. Shown are the two features displaying mathematical singularities, the
critical point (closed circles) and the liquid-gas coexistence line (bold
dashed curve). (b) Same as (a) with the addition of the gas-liquid spinodal
and the Widom line. Along the Widom line, thermodynamic response functions
have extrema in their $T$ dependence. Path $\alpha$ denotes a path along
which the Widom line is crossed, while path $\beta$ denotes a path crossing
the coexistence line. (c) A hypothetical phase diagram for water of possible
relevance to the recent neutron scattering experiments by Chen {\it et al.}
\cite{chenJCP2004, Liu04} on confined water. The negatively sloped
liquid-liquid coexistence line generates a Widom line which extends beyond
the critical point, suggesting that water may exhibit a fragile-to-strong
transition for $P<P_c$ (path $\alpha$), while no dynamic changes will occur
above the critical point (path $\beta$). (d) A sketch of the $P-T$ phase
diagram for the two-scale Jagla model. Upon cooling at constant pressure
above the critical point (path $\alpha$), the liquid changes, as the path
crosses the Widom line, from a low density state (characterized by a
non-glassy Arrhenius dynamics) to a high density state (characterized by
non-Arrhenius dynamics) as the path crosses the Widom line. Upon cooling at
constant pressure below the critical point (path $\beta$), the liquid remains
in the LDL phase as long as path $\beta$ does not cross the LDL spinodal
line. Thus one does not expect any dramatic change in the dynamic behavior
along the path $\beta$.}
\label{P-T-FS}
\end{figure}

\noindent
By definition, in a first order phase transition, thermodynamic state
functions such as density $\rho$ and enthalpy $H$ change discontinuously as
we cool the system along a path crossing the equilibrium coexistence line
[Fig.~\ref{P-T-FS}(a), path $\beta$]. However in a real experiment, this
discontinuous change may not occur at the coexistence line since a substance
can remain in a supercooled metastable phase until a limit of stability (a
spinodal) is reached \cite{pgdbook} [Fig.~\ref{P-T-FS}(b), path $\beta$].

If the system is cooled isobarically along a path above the critical
pressure $P_{c}$ [Fig.~\ref{P-T-FS}(b), path $\alpha$], the state
functions continuously change from the values characteristic of a high
temperature phase (gas) to those characteristic of a low temperature
phase (liquid).  The thermodynamic response functions which are the
derivatives of the state functions with respect to temperature [e.g.,
isobaric heat capacity $C_P=(\partial H/\partial T)_P$] have maxima at
temperatures denoted $T_{\rm max}(P)$. Remarkably these maxima are still
prominent far above the critical pressure \cite{AnisimovbookA, AnisimovbookB, AnisimovbookC,Anisimovbook}, and the values of the
response functions at $T_{\rm max}(P)$ (e.g., $C_{P}^{\rm max}$) diverge
as the critical point is approached. The lines of the maxima for
different response functions asymptotically approach one another as the
critical point is approached, since all response functions become
expressible in terms of the correlation length.  This asymptotic line is
sometimes called the Widom line, and is often regarded as an extension
of the coexistence line into the ``one-phase region.''
 
If the system is cooled at constant pressure $P_0$, and $P_0$ is not too
far from the critical pressure $P_c$, then there are two classes of
behavior possible.  (i) If $P_0>P_c$ (path $\alpha$), then
experimentally-measured quantities will change dramatically but
continuously in the vicinity of the Widom line (with huge fluctuations
as measured by, e.g., $C_P$).  (ii) If $P_0<P_c$ (path $\beta$),
experimentally-measured quantities will change discontinuously if the
coexistence line is actually seen. However the coexistence line can be
difficult to detect in a pure system due to metastability, and changes
will occur only when the spinodal is approached where the gas phase is
no longer stable.  The changes in behavior may include not only static
quantities like response functions \cite{Anisimovbook} but also dynamic
quantities like diffusivity.

\begin{figure}[htb]
\includegraphics[width=5cm,height=5cm]{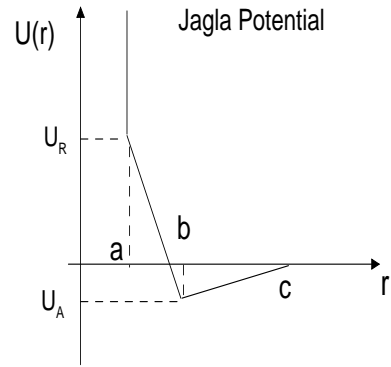}
\caption{The ``two-scale'' Jagla ramp potential with attractive and repulsive
ramps. Here $U_{R}=3.5U_{0}$, $U_{A}=-U_{0}$, $a$ is the hard core diameter,
$b=1.72a$ is the soft core diameter, and $c=3a$ is the long distance
cutoff. In the simulation, we use $a$ as the unit of length, and $U_{0}$ as
the unit of energy.}
\label{fig:jagla-pot}
\end{figure}
\begin{figure*}
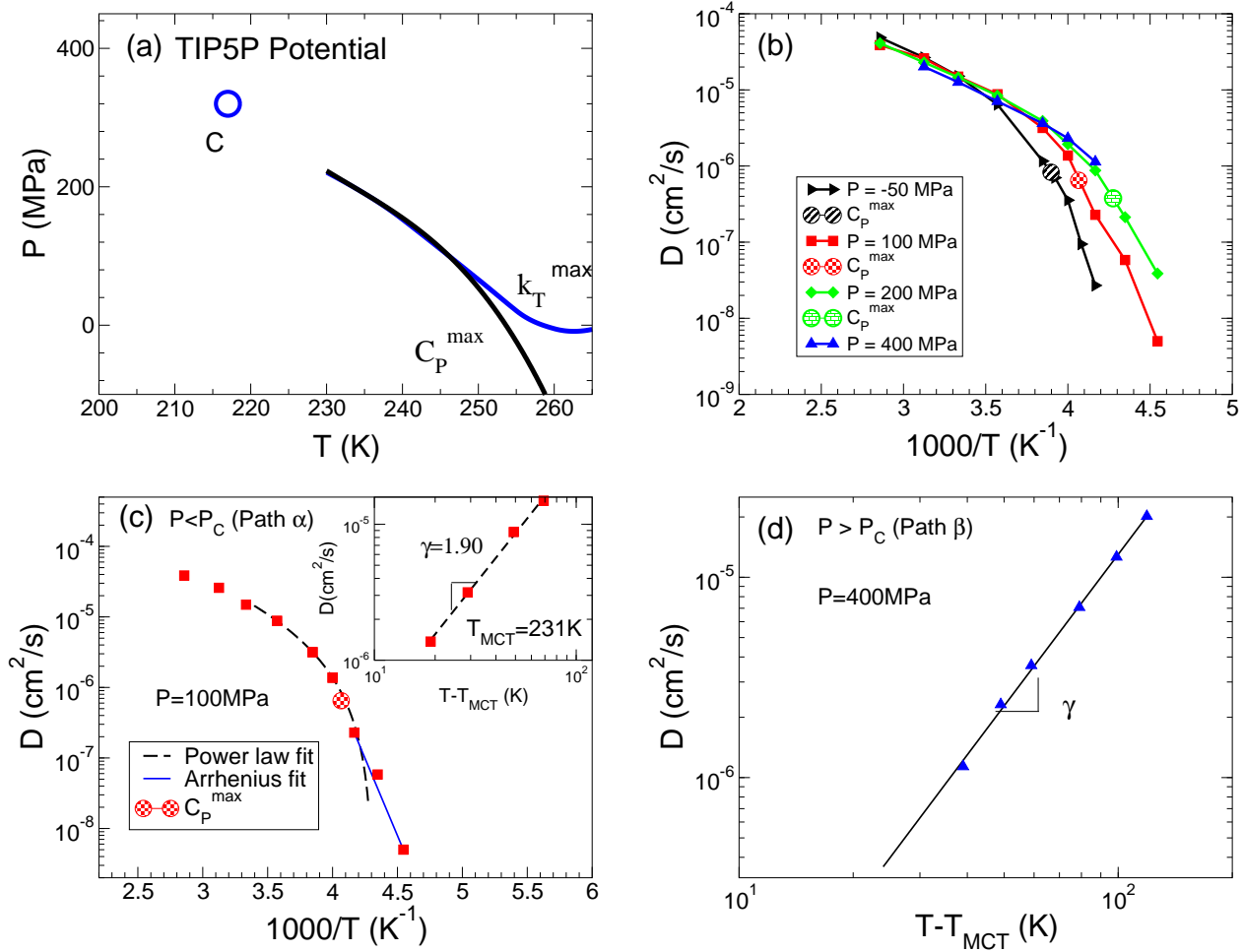

\begin{center}
\includegraphics[width=8cm,angle=0]{fig3a.eps}\hfil
\includegraphics[width=8cm,angle=0]{fig3b.eps}
\\[3ex]
\includegraphics[width=8cm,angle=0]{fig3c.eps}\hfil
\includegraphics[width=8cm,angle=0]{fig3d.eps}
\end{center}
\caption{Results for the TIP5P potential. (a) Relevant part of the phase
diagram, showing the liquid-liquid critical point $C$ at $P_{c} \approx
320$~MPa and $T_{c} \approx 217$~K, the line of isobaric specific heat maxima
$C_P^{\rm max}$ and the line of isothermal compressibility maxima $K_{T}^{\rm
max}$. (b) Arrhenius plot of the diffusion constant $D$ as a function of
$1000/T$ along different isobars. The filled circles indicate the
temperatures at which the $C_P^{\rm max}$ line is crossed. (c) Arrhenius plot
of $D$ as a function of $1000/T$ for $P=100$~MPa (path $\alpha$). At high
temperatures, $D$ can be fit by $D\sim (T-T_{MCT})^\gamma$ (dashed line, also
shown in the inset), where $T_{MCT} \approx 231$~K and $\gamma \approx
1.9$. At low temperatures the dynamic behavior changes to that of a liquid
where $D$ is Arrhenius (solid line). (d) Log-log plot of $D$ as a function of
$T-T_{MCT}$ for $P=400$~MPa (path $\beta$). The behavior of $D$ remains
non-Arrhenius for the entire temperature range and is consistent with $D\sim
(T-T_{MCT})^\gamma$, with $T_{MCT} \approx 201$~K and $\gamma \approx
2.5$. Note that the power law fits for $\gamma$ and $T_{MCT}$ are subject to
error due to the relatively small ranges of $D$ and $T-T_{MCT}$.}
\label{fig:tip5p}
\end{figure*}

\begin{figure*}
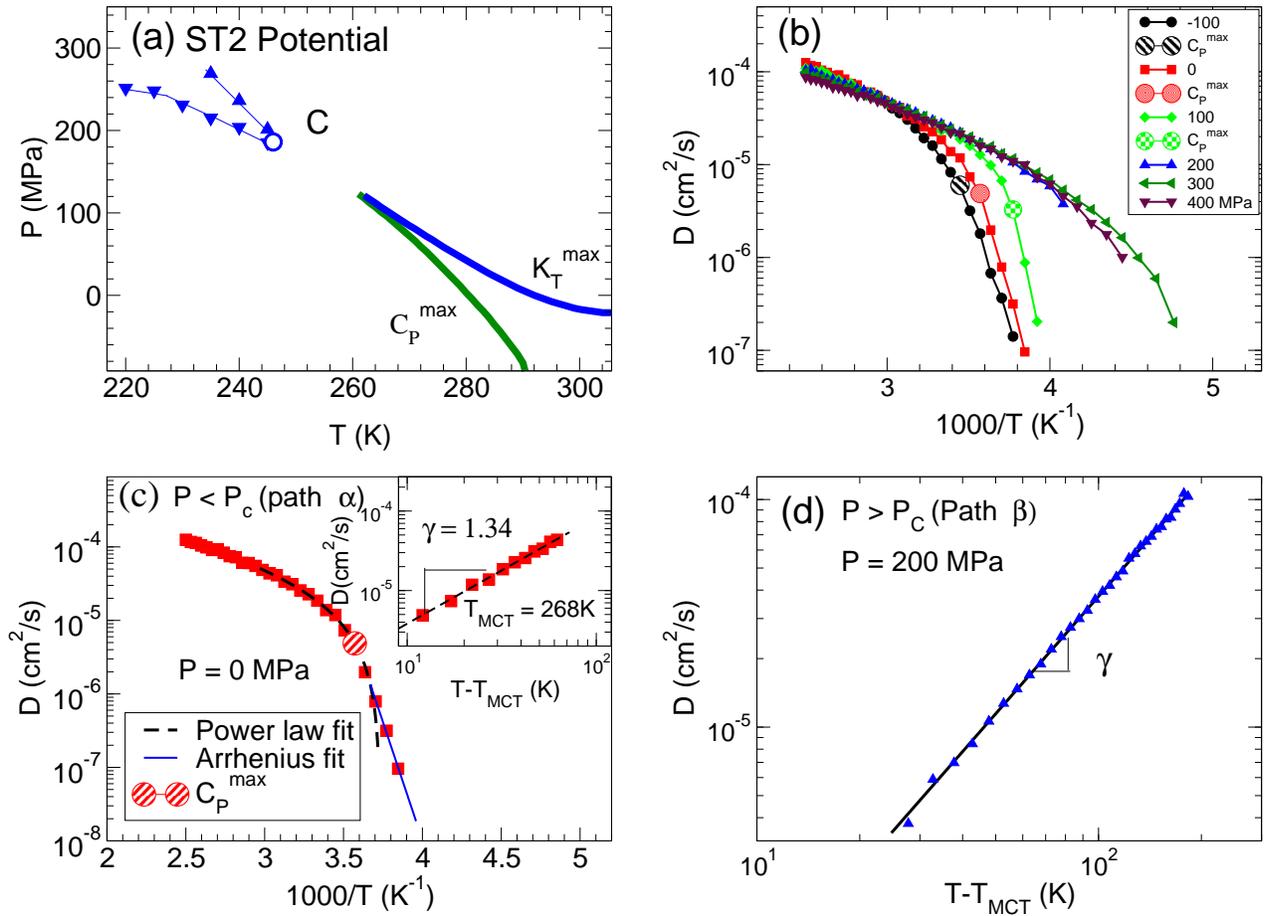

\begin{center}
\includegraphics[width=8cm]{fig4a.eps}\hfil
\includegraphics[width=8cm]{fig4b.eps}
\\[3ex]
\includegraphics[width=8cm]{fig4c.eps}\hfil
\includegraphics[width=8cm]{fig4d.eps}
\end{center}
\caption{Analog of Fig.~\ref{fig:tip5p} for the ST2 potential. (a) Relevant
part of the phase diagram, showing the liquid-liquid critical point $C$ at
$P_{c} \approx 186$~MPa and $T_{c} \approx 246$~K, the line of isobaric
specific heat maxima $C_P^{\rm max}$, the line of isothermal compressibility
maxima $K_{T}^{\rm max}$, and the spinodal lines. (b) Arrhenius plot of the
diffusion constant $D$ as a function of $1000/T$ along different isobars. The
filled circles indicate the temperatures at which the $C_P^{\rm max}$ line is
crossed. (c) Arrhenius plot of $D$ as a function of $1000/T$ for $P=0$~MPa
(path $\alpha$). At high temperatures, $D$ can be fit by $D\sim
(T-T_{MCT})^\gamma$ (dashed line, also shown in the inset) where $T_{MCT}
\approx 268$~K and $\gamma \approx 1.34$. At low temperatures the dynamic
behavior changes to that of a liquid where $D$ is Arrhenius (solid line). (d)
Log-log plot of $D$ as a function of $T-T_{MCT}$ for $P=200$~MPa (path
$\beta$). The behavior of $D$ remains non-Arrhenius for the entire
temperature range and is consistent with $D\sim (T-T_{MCT})^\gamma$, with
$T_{MCT} \approx 217$~K and $\gamma \approx 1.7$. Note that the power law
fits for $\gamma$ and $T-T_{MCT}$ are subject to error due to the relatively
small ranges of $D$ and $T-T_{MCT}$.}
\label{fig:st2}
\end{figure*}

\begin{figure*}[htb]
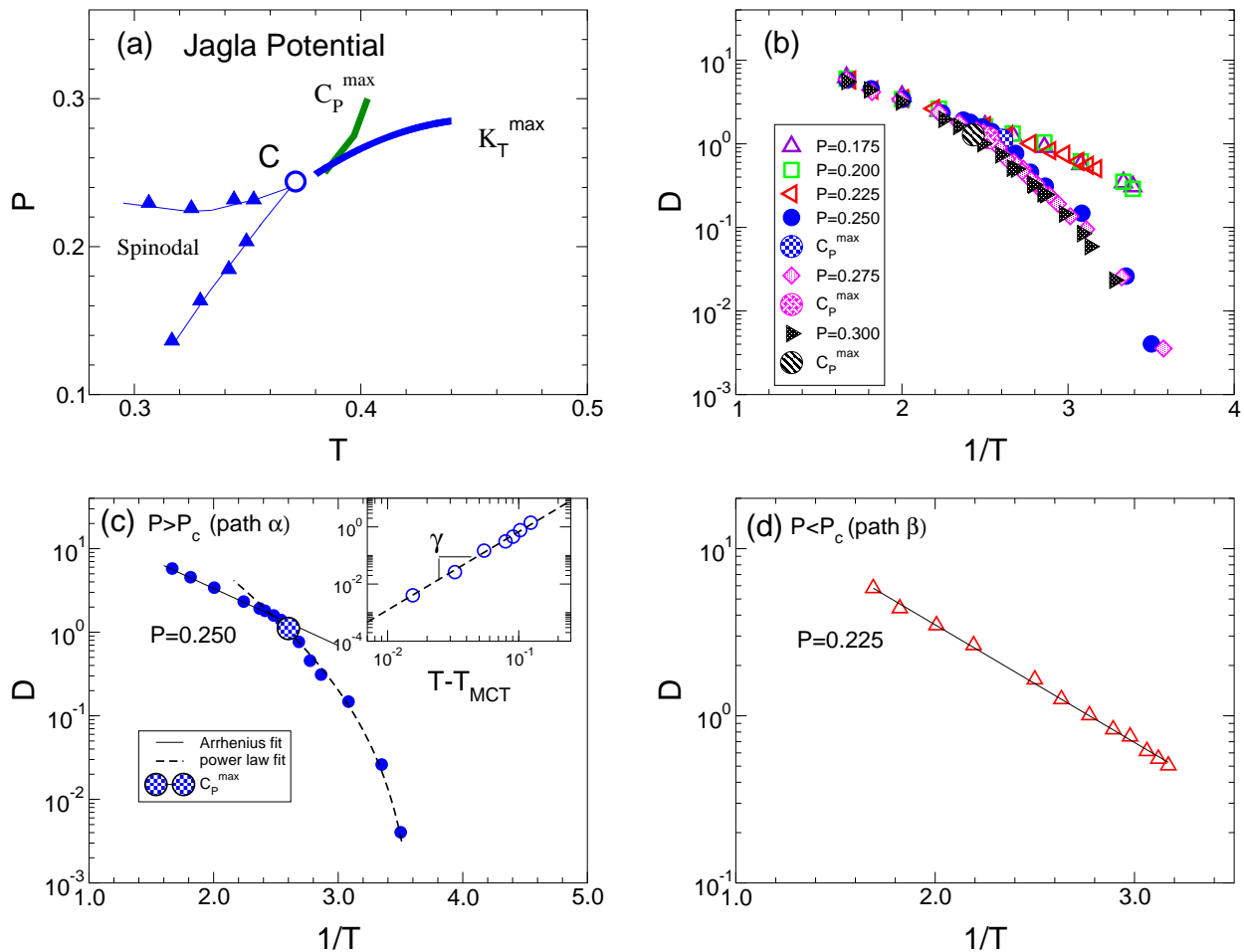

\begin{center}
\includegraphics[width=8cm]{fig5a.eps}\hfil
\includegraphics[width=8cm]{fig5b.eps}
\\[3ex]
\includegraphics[width=8cm]{fig5c.eps}\hfil
\includegraphics[width=8cm]{fig5d.eps}
\end{center}
\caption{Analog of Figs.~\ref{fig:tip5p} and ~\ref{fig:st2} for the two-scale
 Jagla potential. (a) Phase diagram in the vicinity of the liquid-liquid
 critical point $C$ located at $P_{c} \approx 0.24$ and $T_{c} \approx 0.37$,
 the line of isobaric specific heat maxima $C_{P}^{\rm max}$, the line of
 isothermal compressibility maxima $K_{T}^{\rm max}$, and the spinodal
 lines. (b) The $T$-dependence of diffusivity along constant pressure
 paths. Several paths $\alpha$ and paths $\beta$ are shown: (i)
 $P=0.175,0.200,0.225 < P_{c}$ (paths $\beta$ in Fig.~\ref{P-T-FS}(d), along
 which the system remains in the LDL phase). (ii) $P=0.250, 0.275, 0.30
 >P_{c}$ (paths $\alpha$ in Fig.~\ref{P-T-FS}(d), along which the system does
 not remain in the LDL-like state, but the dynamic behavior changes from
 Arrhenius to non-Arrhenius). (c) $D$ as a function of $1/T$ for $P=0.250$
 (path $\alpha$). At high temperatures, the fit is Arrhenius
 $D\sim\exp(-1.59/T)$ (solid line), while at low temperatures the results are
 consistent with $D\sim (T-T_{MCT})^\gamma$ with $T_{MCT} \approx 0.27$ and
 $\gamma \approx 2.7$ (dashed line, also shown in the inset). (d) For
 $P=0.225$ (path $\beta$), $D$ is Arrhenius for the entire temperature range
 and can be fit by $D\sim\exp(-1.62/T)$. The unit of $D$ is $a\sqrt{U_{0}/m}$ and the unit of $P$ is $U_{0}/a^{3}$.}
\label{fig:D-T-P}
\end{figure*}
In the case of water---the most important solvent for biological function
\cite{Robinson1996,Bellisent}---a significant change in dynamical properties
has been suggested to take place in deeply supercooled states
\cite{Angell93,pabloreview,Francisphysica2003,angellreview}. Unlike other
network forming materials \cite{Kob99}, water behaves as a fragile liquid in
the experimentally accessible window
\cite{prielmeierXX,pabloreview,LudemanXX}.  Based on analogies with other
network forming liquids and with the thermodynamic properties of the
amorphous forms of water, it has been suggested that, at ambient pressure,
liquid water should show a crossover between fragile behavior at high T to
strong behavior at low T \cite{Angell93, Itonature1999,Jagla99a, Jagla99b,
Jagla99c,Tanaka03} in the deep supercooled region of the phase diagram below
the homogeneous nucleation line. This region may contain the hypothesized
liquid-liquid critical point \cite{poole1}, the terminal point of a line of
first order liquid-liquid phase transitions. According to one current
hypothesis, the liquid-liquid critical point is the thermodynamic source of
all water's anomalies
\cite{poole1,Mishima1998nature,franzese2001nature,lanaveprl}. This region has
been called the ``no-man's land'' because to date no experiments have been
able to make direct measurements on the {\it bulk\/} liquid phase
\cite{Mishima1998nature}. Recently the fragility transition in confined water
was studied experimentally \cite{Bergman00,chenJCP2004,Liu04} since
nucleation can be avoided in confined geometries. Also, a dynamic crossover
has been associated with the liquid-liquid phase transition in silicon and
silica \cite{Poolenature2001,sastrynature2003}. In this work, we offer an
interpretation of the dynamic crossover (called a fragility transition or
fragile-strong transition by many authors) in water as arising from crossing
the Widom line emanating from the hypothesized liquid-liquid critical point
\cite{Poolenature2001} [Fig.~\ref{P-T-FS}(c), path $\alpha$]. Our
thermodynamic and structural interpretation of the dynamic crossover may not
hold for liquids for which the fragile-strong dynamic crossover can be caused
by other mechanisms, as discussed in \cite{GC}. 
\section{Methods} 
Using molecular dynamic (MD) simulations, we study three
models, each of which has a liquid-liquid critical point. Two of the models,
(the TIP5P \cite{JorgensenXX} and the ST2 \cite{StillingerXX}) treat the
water molecule as a multiple-site rigid body, interacting via electrostatic
site-site interactions complemented by a Lennard-Jones potential. The third
model is the spherical ``two-scale'' Jagla potential with attractive and
repulsive ramps [Fig.~\ref{fig:jagla-pot}] which has been studied in the
context of liquid-liquid phase transitions and liquid anomalies
\cite{Jagla99a, pradeepPRE}. For all three models, we evaluate the loci of
maxima of the relevant response functions, compressibility and specific heat,
which coincide close to the critical point and give rise to the Widom
line. We provide evidence that, for all three potentials, a dynamic crossover
occurs when the Widom line is crossed.

Our results for the TIP5P potential are based on MD simulations of a system
of $N=512$ molecules, carried out both in the NPT and NVT ensembles using the
techniques described in \cite{YamadaXX}. For ST2 simulations $N=1728$
molecules are used and all the simulations are carried out in NVT
ensemble. For the Jagla potential, discrete molecular dynamics simulation
\cite{pradeepPRE} implemented for $N=1728$ particles interacting with step
potentials \cite{Buldyrev2003physicA} is used in both NVT and NVE
ensembles.
\section{Results}

Fig.~\ref{fig:tip5p}(a) shows for TIP5P the relevant portion of the $P-T$
phase diagram. A liquid-liquid critical point is observed
\cite{YamadaXX,Paschek05}, from which the Widom line develops. The
coexistence curve is negatively sloped, so the Clapeyron equation implies
that the high-temperature phase is a high-density liquid (HDL) and the
low-temperature phase is a low-density liquid (LDL). Fig.~\ref{fig:tip5p}(b)
shows the $T$ dependence of the diffusion coefficient $D$, evaluated from the
long time limit of the mean square displacement along isobars. The isobars
crossing the Widom line [Fig.~\ref{fig:tip5p}(c), path $\alpha$] show a clear
crossover from a non-Arrhenius behavior at high $T$ [which can be well
fit by a power law function $D \sim (T-T_{\rm MCT})^{\gamma}$], consistent
with the mode coupling theory predictions \cite{gotze2}), to an
Arrhenius behavior at low $T$ [which can be described by
$D\sim\exp(-E_{a}/T)$]. The crossover between these two functional forms
takes place when crossing the Widom line.

For paths $\beta$ [Fig.~\ref{fig:tip5p}(d)], crystallization occurs in TIP5P
\cite{YamadaXX}, so the hypothesis that there is no fragility transition
cannot be checked at low temperature. Hence we consider a related potential,
ST2, for which crystallization is absent within the time scale of the
simulation. Simulation details are described in \cite{poole2}. This potential
also displays a liquid-liquid critical point \cite{poole1,poole2}, as seen in
the phase diagram of Fig.~\ref{fig:st2}(a). The analog of
Fig.~\ref{fig:tip5p}(b) is shown in Fig.~\ref{fig:st2}(b). We confirm that
along paths $\alpha$ a fragility transition takes place
[Fig.~\ref{fig:st2}(c)]. Moreover, along paths $\beta$ the $T$ dependence of
$D$ does not show any sign of crossover to Arrhenius behavior and the fragile
behavior is retained down to the lowest studied temperature (note that
$10^{3}/T$ extends to $4.8K^{-1})$. Indeed, for paths $\beta$, the entire $T$
dependence can be fit by a power law $(T-T_{\rm MCT})^{\gamma}$
[Fig.~\ref{fig:st2}(c)].

Thus we see that the simulations for both TIP5P and ST2 water models support
the connection between the Widom line and the dynamic fragility
transition. It is natural to ask which features of the water molecular
potential are responsible for the properties of water discussed here,
especially because water's unusual properties are shared by several other
liquids whose inter-molecular potential has two energy (length) scales
\cite{sastrynature2003,Poolenature2001}. We next investigate the two-scale
spherically symmetric Jagla potential. The Jagla model displays---without the
need to supercool---a liquid-liquid coexistence line which, unlike water, has
a positive slope, implying that the Widom line is now crossed along $\alpha$
paths with $P>P_c$ [Figs.~\ref{P-T-FS}(d) and \ref{fig:D-T-P}(a)]. There is a
crossover in the behavior of $D(T)$ when the $C_{P}^{\rm max}$ line is
crossed [Figs.~\ref{fig:D-T-P}(b) and~\ref{fig:D-T-P}(c)]. At high
temperature, $D$ exhibits an Arrhenius behavior [Figs.~\ref{fig:D-T-P}(b) and
~\ref{fig:D-T-P}(c)], while at low temperature it follows a non-Arrhenius
behavior, consistent with a power law. Along a $\beta$ path ($P<P_c$), $D(T)$
follows the Arrhenius behavior over the entire studied temperature range
[Figs.~\ref{fig:D-T-P}(b) and~\ref{fig:D-T-P}(d)]. Thus, the dynamic
crossover coincides with the location of the $C_{P}^{\rm max}$ line,
extending the conclusion of the TIP5P and ST2 potentials to a general
two-scale spherically symmetric potential.
\section{ Discussion and Summary}
Before concluding, we note that our findings are consistent with the
possibility that the observed dynamic crossover along path $\alpha$ is
related to the behavior of $C_P$, suggesting that enthalpy or entropy
fluctuations may have a strong influence on the dynamic properties. The role
of $C_P$ is consistent with expectations based on the Adam-Gibbs \cite{AG}
interpretation of the water dynamics \cite{oldangel,scalanature2000} and of
the fragility transition \cite{Poolenature2001,Francisphysica2003}.

For both water and the Jagla model, crossing the Widom line is
associated with a change in the T-dependence of the dynamics. In the
case of water, $D(T)$ changes from non-Arrhenius (``fragile'') to
Arrhenius (``strong'') behavior, while the structural and thermodynamic
properties change from those of HDL to those of LDL. For the Jagla
potential, due to the {\it positive\/} slope of the Widom line, $D(T)$
changes from Arrhenius to non-Arrhenius while the structural and
thermodynamic properties change from those of LDL to those of HDL.

In summary, our results for water are consistent with the experimental
observation in confined water of (i) a fragility transition for
$P<P_{c}$ \cite{chenJCP2004,Liu04}, and (ii) a peak in $C_{P}$ upon
cooling water at atmospheric pressure \cite{Oguni}. Thus our work offers
a plausible interpretation of the results of Ref.~\cite{Liu04}
consistent with the existence of a liquid-liquid critical point located
in the no-man's land.

We thank C. A. Angell, G. Franzese, J. M. H. Levelt Sengers, L. Liu,
M. Mazza, S. Sastry, F. W. Starr, B. Widom, and Z. Yan for helpful
discussions and NSF grant CHE~0096892 and MIUR-FIRB for support. We thank
D. Chandler and J. P. Garrahan for helpful criticisms on the manuscript. We
also thank the Boston University Computation Center, Yeshiva University, and
StFX hpcLAB (high performance computing laboratory) for allocation of CPU
time.


\end{document}